\documentclass[aps,pre,twocolumn,groupedaddress]{revtex4}
\usepackage{color}
\usepackage{tabularx}
\usepackage{epsfig}
\usepackage{amsmath}
\usepackage{graphicx}

\begin{document}

\title{Evidence for non-universal scaling in dimension four Ising spin glasses}

\author{P. H.~Lundow} 

\affiliation {Department of Mathematics and Mathematical Statistics,
  Ume{\aa} University, SE-901 87, Sweden}

\author{I. A.~Campbell}

\affiliation{Laboratoire Charles Coulomb (L2C), UMR 5221
  CNRS-Université de Montpellier, Montpellier, F-France
}

\date{\today}


\begin{abstract}
The critical behavior of the Binder cumulant for Ising spin glasses in
dimension four are studied through simulation measurements. Data for
the bimodal interaction model are compared with those for the
Laplacian interaction model.  Special attention is paid to scaling
corrections. The limiting infinite size value at criticality for this
dimensionless variable is a parameter characteristic of a universality
class.  This critical limit is estimated to be equal to $0.523(3)$ in
the bimodal model and to $0.473(3)$ in the Laplacian model.
\end{abstract}

\maketitle 



For standard second order transitions, Renormalization Group Theory
(RGT) provides an elegant and detailed explanation of
universality. Thus in the family of simple ferromagnets, within a
universality class of models having space dimension $d$ and spin
dimensionality $n$, all models have identical critical properties
corresponding to an isolated fixed point in the renormalization group
flow.  The only documented exceptions appear all to be cases of
specific spin models in dimension two (discussed for instance in
Ref.~\cite{cardy:87}); for these models the critical behavior is more complicated,
with critical exponents varying continuously when a control parameter
is modified.  The corresponding renormalization group scenario
consists of a line of fixed points rather than an isolated fixed
point, with motion along the line produced by a marginal operator.


The Ising spin glasses (ISGs) which we will consider have symmetric
(positive and negative) random near neighbor interactions rather than
the regular interactions with fixed sign of a simple ferromagnet; the
theoretical situation for critical behavior in ISGs is far less
advanced than for the standard models. The ISG upper critical
dimension is known to be six, but it was found thirty years ago that
the $\epsilon$-expansions in ISGs are not fully predictive since the
first few orders have a non-convergent behavior and higher orders are
not known~\cite{gardener:84}. This can be taken as an indication that
a fundamentally different theoretical approach is required for RGT at
spin glass transitions, and indeed \lq\lq classical tools of RGT
analysis are not suitable for spin glasses\rq\rq{}
\cite{parisi:01,castellana:11,angelini:13} although no explicit
theoretical predictions have been made so far concerning the important
question of universality in these systems.


Claims of universality in ISGs have been made repeatedly based on
numerical
data~\cite{bhatt:88,katzgraber:06,hasenbusch:08,baity:13,jorg:08a}. Here,
from a detailed analysis of numerical simulation measurements on ISGs
in dimension four we come to the empirical conclusion that, on the
contrary, the critical properties of these systems depend on the form
of the interaction distribution. A breakdown of universality at a
continuous spin glass transition for a dimension well above two may be
a symptom of the need for a novel RGT approach in this class of
models.

The ISG Hamiltonian is 
\begin{equation}\label{ham}
  \mathcal{H}= - \sum_{ij}J_{ij}S_{i}S_{j}
\end{equation} with the near
neighbor symmetric distributions normalized to $\langle
J_{ij}^2\rangle=1$.  We use the inverse temperature $\beta=1/T$ as
thermal parameter. The Ising spins sit on simple hyper-cubic lattices
with periodic boundary conditions. The spin overlap parameter is
defined by
\begin{equation}\label{qdef} 
  q = \frac{1}{L^d}\sum_{i}S_{i}^{A} S_{i}^{B}
\end{equation} 
where $A$ and $B$ indicate two copies of the same system.
We have studied in dimension $4$ the bimodal model with a $\pm J$
interaction distribution, and the Laplacian model with a $P(J_{ij})
\sim \exp(-|J_{ij}|)$ interaction distribution. 

Simulations in ISGs are very much more laborious than the equivalent
simulations in simple ferromagnets because equilibration is slow and
averages must be taken over large numbers of samples. The simulations
were carried out using the exchange Monte-Carlo
method~\cite{hukushima:96} for equilibration using so called
multi-spin coding. In the bimodal model measurements were made on
$2^{14}$ individual samples (or $J_{ij}$-realizations) for $3\le L\le
7$, on $2^{13}$ samples for $8\le L\le 12$, and on $2^{12}$ samples
for $L=13$ and $L=14$. For the Laplacian model, measurements were made
on $2^{13}$ samples for $3\le L\le 12$. After every sweep an exchange
was attempted with a success rate of at least 30\%. At least 40
temperatures were used forming a geometric progression reaching down
to $\beta_{\max}=0.55$ in the bimodal case and $\beta_{\max}=0.70$ in
the Laplacian case.

This ensures that our data span the critical temperature region which
is essential for the FSS fits. Near the critical temperature the
$\beta$ step length was at most $0.03$. The various systems were
deemed to have reached equilibrium when the sample average
susceptibility for the lowest temperature showed no trend between
runs. For example, in the Laplacian case for $L=12$ this means about
$200 000$ sweep-exchange steps.

After equilibration, at least $200 000$ measurements were made for
each sample for all sizes, taking place after every sweep-exchange
step.  We registered the energy $E(\beta,L)$, the correlation length
$\xi(\beta,L)$, the spin overlap moments $\langle |q| \rangle$,
$\langle q^2\rangle$, $\langle |q|^3\rangle$, $\langle q^4\rangle$ and
the corresponding link overlap $q_{\ell}$ moments, where the link
overlap is defined as
\begin{equation}\label{qldef} 
  q_{\ell} = \frac{1}{dL^d}\sum_{ij}S_{i}^{A}S_{j}^{A}S_{i}^{B}S_{j}^{B}
\end{equation}
In addition, some correlations $\langle E(\beta,L),U(\beta,L)\rangle$
between the energy and observables $U(\beta,L)$ were also registered
so that thermodynamic derivatives could be evaluated using the
relation $\partial U(\beta,L)/\partial \beta = \langle U(\beta,L),
E(\beta,L)\rangle-\langle U(\beta,L) \rangle\langle
E(\beta,L)\rangle$, see e.g. Ref.~\cite{ferrenberg:91}.  Bootstrap
analyses of the errors in the derivatives as well as in the
observables $U(\beta,L)$ themselves were carried out.

In Ref.~\cite{jorg:08} J\"{o}rg and Katzgraber used an elegant scaling
display of raw numerical data to test for universality in Ising Spin
Glasses (ISGs). They plot the ratio $y(\beta,L) =
g(\beta,2L)/g(\beta,L)$ against $x(\beta,L) = g(\beta,L)$ where
\begin{equation}\label{gdef}
  g(\beta,L)=\frac{1}{2}\left(3-\frac{[\langle q^4\rangle]}{[\langle q^2\rangle]^2}\right)
\end{equation}
is the Binder cumulant for inverse temperature $\beta$ and lattice
size $L$, with $q$ the spin glass order parameter of Eq.~\eqref{qdef}
and $[\cdots]$ denoting the average taken over the samples. They
studied numerically two ISGs in dimension $4$, one with a Gaussian
interaction distribution and one with a diluted bimodal distribution.
Over the range of temperatures used for the measurements, which
extended well into the ordered phase, the scaled data points were
independent of $L$ and followed the same curve for the two systems to
within the statistics.  J\"org and Katzgraber concluded that these
results were evidence of universality in ISGs.

\begin{figure}
  \includegraphics[width=3.5in]{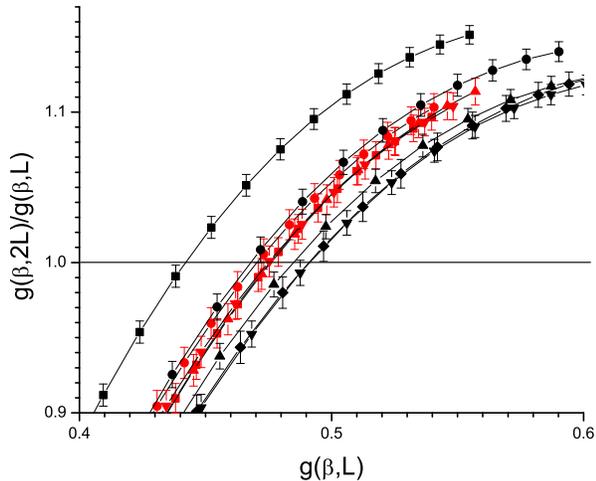}
  \vskip -5ex
  \caption{(Color online) Scaling plots for the Binder cumulant of
    $4$d ISGs, $g(\beta,2L)/g(\beta,L)$ vs $g(\beta,L)$.  Red:
    Laplacian interactions, black: bimodal interactions. Square,
    circle, triangle, inverted triangle, diamond symbols for
    $L=3,4,5,6,7$ respectively.}\protect\label{fig:1}
\end{figure}

In Fig.~\ref{fig:1} we show the same scaling plot as that of
Ref.~\cite{jorg:08} in dimension 4 but using instead standard bimodal
interactions and compare them to Laplacian interactions. The
temperatures span the critical temperatures.
 
For the Laplacian ISG, our data show scaling with no correction term
to within the statistics; the scaling curves are almost
indistinguishable from those for the models of
Ref.~\cite{jorg:08}. The bimodal data on the other hand show a strong
$L$ dependence due to large finite size scaling corrections; the
scaling curve $y(x)$ moves continously to the right with increasing
$L$. With a natural extrapolation the thermodynamic (large $L$) limit
scaling curve for the bimodal interaction ISG will lie well to the
right of the $L$-independent Laplacian curve, so the two models appear
not to be in the same universality class.

Standard finite size scaling expressions which include a single
leading conformal correction term lead to a size dependence $\beta_c -
\beta_{\mathrm{cross}}(L) = A L^{-(\omega+1/\nu)}$ where
$\beta_{\mathrm{cross}}$ is the crossing point where
$g(\beta,2L)=g(\beta,L)$ (represented by $y(x)=1$ in
Fig.~\ref{fig:1}), where $\omega$ is the correction to scaling
exponent. In the dimension $4$ bimodal ISG, $\omega$ has been
estimated by simulations to be $1.04(10)$ \cite{banos:12}.  From high
temperature series expansion (HTSE) measurements $\theta = \omega\nu
\approx 1.5$ \cite{daboul:04} so $\omega \approx 1.3$. A natural
extrapolation of the present bimodal data to infinite $L$ assuming
$\omega \approx 1.2$ gives a thermodynamic limit estimate which is
certainly considerably larger than the Laplacian crossing point limit.

\begin{figure}
  \includegraphics[width=3.5in]{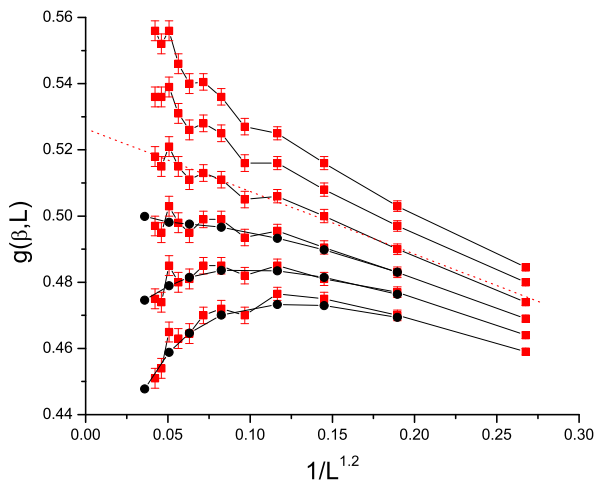}
  \vskip -5ex
  \caption{(Color online) The Binder cumulant of the $4$d bimodal ISG
    near criticality. Inverse temperatures $\beta = 0.510, 0.5075,
    0.505, 0.5025, 0.500. 0.4975$ from top to bottom. Red squares :
    present data. Black circles : read from Ref.~\cite{banos:12}. The
    two data sets are consistent. The dashed straight line indicates
    criticality.}\protect\label{fig:2}
\end{figure}

\begin{figure}
  \includegraphics[width=3.5in]{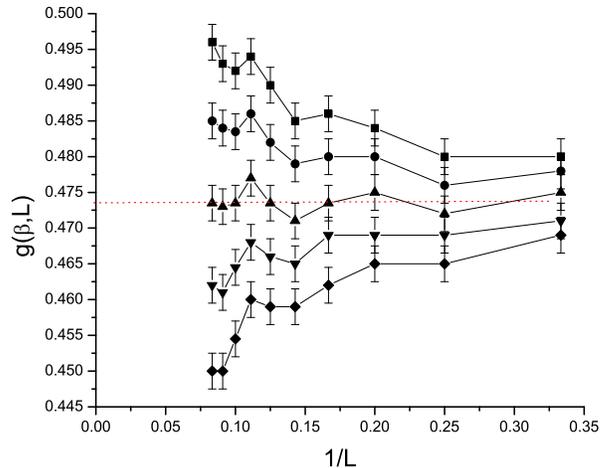}
  \vskip -5ex
  \caption{(Color online) The Binder cumulant of the $4$d Laplacian
    ISG near criticality.  Inverse temperatures $\beta = 0.626, 0.624,
    0.622, 0.620, 0.618$ from top to bottom.  The dashed straight line
    indicates criticality. }\protect\label{fig:3}
\end{figure}

Data near criticality for the bimodal and Laplacian ISGs are shown in
a different form in Fig.~\ref{fig:2} and Fig.~\ref{fig:3}
respectively.  Near criticality
\begin{equation}
g(\beta,L) = g_{c} + AL^{-\omega} + B(\beta-\beta_{c})L^{1/\nu}
\end{equation}
The bimodal data are consistent with $\beta_{c}= 0.505(1)$, $\omega
\approx 1.2$ and $g_{c}=0.523(3)$. The Laplacian data are consistent
with $\beta_{c}= 0.622(1)$, $g_{c}=0.473(3)$ and a negligible
correction.  The $g_{c}$ values estimated for the Gaussian and dilute
bimodal models in Ref.~\cite{jorg:08a} are $0.470(5)$ and $0.472(2)$
which are similar to the Laplacian value.

\begin{figure}
  \includegraphics[width=3.5in]{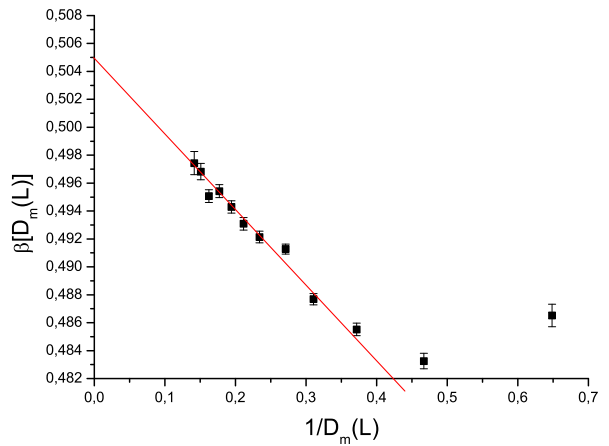}
  \vskip -5ex
  \caption{(Color online) Scaling plot for the bimodal model
    $W_{q}(\beta,L)$ derivative peak location $\beta_m$ against the
    inverse derivative peak height $1/D_m(L)$.} \protect\label{fig:4}
\end{figure}

The bimodal $\beta_{c}$ value is confirmed independently by
thermodynamic derivative data on dimensionless observables
$U(\beta,L)$. The $U(\beta,L)$ curve becomes steeper and steeper with
increasing $L$, and tends to a step function centered on $\beta_{c}$
in the large $L$ limit. Calling the peak in the derivative $D_m(L)
=[\partial U(\beta,L)/\partial \beta]_{\max}$ and its location (the
pseudo critical temperature) $\beta_m$, then the inverse of the
derivative peak height $x(L)=1/D_{m}(L)$ and the corresponding inverse
temperature location shift $\beta_c-\beta_m$ both scale as
$L^{-1/\nu}[1+ aL^{-\omega}] $ \cite{ferrenberg:91}. So at large $L$,
the points $y(L) =\beta_m(L)$ plotted against $x(L)$ extrapolate
linearly to $y(\infty) = \beta_{c}$ at $x(\infty)=0$. An example of
this type of plot with the dimensionless observable
\begin{equation}
  W_{q}(\beta,L) = \frac{1}{\pi-2} \left(\pi\frac{\lbrack\langle
    |q|\rangle\rbrack^2}{\lbrack\langle q^2\rangle\rbrack} - 2\right)
  \label{Wqdef}
\end{equation}
is shown for the bimodal model in Fig.~\ref{fig:4}. From such plots an
independent estimate $\beta_{c}= 0.505(1)$ is obtained for the bimodal
model in $4$d~\cite{lundow}.


For the $g(\beta_{c},L)$ bimodal values to extrapolate finally to a
limiting $g_{c}$ value at infinite $L$ consistent with that of the
Laplacian model would require putative bimodal ISG data for very large
$L$ (data inaccessible with current numerical resources) to bend back
to the left in Fig.~\ref{fig:1} or to sharply bend down in
Fig.~\ref{fig:2} (in an unlikely looking way), instead of
extrapolating in a natural way to the large $L$ critical limit
estimated above.  A necessary condition for this ``backbending'' is
the presence of a hypothetical further correction term which begins to
influence the data only at $L > 14$, and so has an extremely small
exponent (and a prefactor $A$ of the opposite sign).

We can search for potential candidate terms for the hypothetical
backbending.  In addition to the conformal correction in principle
there can also be an analytic correction. 
This term would have an exponent $\omega_a\approx 2$, as in
\cite{ballesteros:99} where in the site percolation context \lq\lq the
subleading analytical corrections for most operators go as
$L^{-\gamma/\nu}\approx L^{-2}$\rq\rq, so such an analytic correction
cannot play the role of the hypothetical very small exponent term.
Turning back to the conformal corrections, the first term in the RGT
$\epsilon$-expansion for the ISG leading irrelevant operator exponent
is $\theta(d) = 6-d$ \cite{dedominicis,bray}, see
Ref.~\cite{ballesteros:97} for the analogous site percolation
$\epsilon$-expansion. Leading $\epsilon$-expansions terms in ISGs give
useful qualitative indications for other critical exponents, and it
turns out that the $\epsilon$-expansion values for $\theta(d)$ :
$\theta(5) \sim 1$, $\theta(4) \sim 2$, $\theta(3) \sim 3$, are
qualitatively consistent with published effective $\theta(d)$ and
$\omega(d) =\theta(d)/\nu(d)$ values from simulations, and from quite
independent HTSE results \cite{klein:91,daboul:04}.  Bimodal ISG
finite size scaling (FSS) estimates in $3$d are $\omega(3) = 1.12(10)$
and $\nu(3)=2.56(4)$ \cite{baity:13}, so $\theta(3) = \omega\nu
\approx 3$. We have seen that in $4$d FSS estimates are
\cite{banos:12} $\theta(4) \approx 1.15$ or $\theta(4) \approx 1.35$
\cite{lundow}. From FSS data for different $5$d ISG models $\omega(5)
\approx 1$ and $\nu(5) \approx 0.75$ \cite{lundow} so $\theta(5)
\approx 1$. HTSE estimates in $4$d ISG models are $\theta(4) \approx
1.4$ \cite{daboul:04} and in $5$d $\theta(5) \approx 1.0$
\cite{klein:91,daboul:04}.

These estimates are all broadly compatible with $\theta(d) \approx
6-d$.  Even though it is hard to pin down an exact value for
$\omega(4)$, consistency definitively excludes a hypothetical leading
conformal correction term in the $4$d bimodal ISG having an exponent
$\omega_{b}(4)$ much smaller than $1$. A correction term with $\omega
\approx 1.2$ for the bimodal ISG can be confidently identified with
the leading conformal correction. By definition, no conformal
correction term with a smaller exponent exists. It can be concluded
that there is no backbending correction, and that the natural
extrapolations of the bimodal model data to the large $L$ limit with
$\omega \approx 1.2$ are valid.


Systems in the same universality class must have identical values for
the infinite size critical limit of a dimensionless parameter such as
the Binder cumulant $g_{c}$. The observation of a critical limit for
the bimodal ISG which is very different from those of the other three
models disproves universality in these $4$d ISGs.

From the existing data there appear to be two possible scenarios : two
classes of ISGs (such as models with continuous distributions and
those with discrete distributions) or alternatively ISG exponents
which vary continuously with a parameter such as the kurtosis of the
interaction distribution.  It would be of interest for statistical
physics in general to obtain further information on the question.
Claims of universality for ISGs in other dimensions should be
re-examined critically.

\begin{acknowledgments}
  The simulations were performed on resources provided by the Swedish
  National Infrastructure for Computing (SNIC) at High Performance
  Computing Center North (HPC2N) and at Chalmers Centre for
  Computational Science and Engineering (C3SE).
\end{acknowledgments}


\begin{thebibliography}{22}
\bibitem{cardy:87} J. L. Cardy, J. Phys. A Math. Gen., {\bf 20} L891 (1987).
\bibitem{gardener:84} E. Gardner, J. Phys. {\bf 45}, 1755 (1984).
\bibitem{angelini:13} M. C. Angelini,  G. Parisi, and F. Ricci-Tersenghi, Phys. Rev. B {\bf 87}, 134201 (2013).
\bibitem{parisi:01} G. Parisi, R. Petronzio, and F. Rosati, Eur. Phys. J. B {\bf 21}, 605 (2001).
\bibitem{castellana:11} M. Castellana, Eur. Phys. Lett. {\bf 95}, 47014 (2011).
\bibitem{katzgraber:06} H. G. Katzgraber, M. Korner, and A. P. Young, Phys. Rev. B {\bf 73}, 224432 (2006).
\bibitem{hasenbusch:08} M. Hasenbusch, A. Pelissetto, and E. Vicari, Phys. Rev. B {\bf 78}, 214205 (2008).
\bibitem{baity:13} M. Baity-Jesi {\it et al.} Phys. Rev. B {\bf 88}, 224416 (2013).
\bibitem{bhatt:88} R. N. Bhatt and A. P. Young, Phys. Rev. B {\bf 37}, 3707 (1988).
\bibitem{jorg:08a} T. J\"{o}rg and H. G. Katzgraber, Phys. Rev. B {\bf 77}, 214426 (2008).
\bibitem{hukushima:96} K. Hukushima and K. Nemoto, J. Phys. Soc. Japan {\bf 65}, 1604 (1996).
\bibitem{ferrenberg:91} A. M. Ferrenberg and D. P. Landau, Phys. Rev. B {\bf 41}, 5081 (1991).
\bibitem{jorg:08} T. J\"{o}rg and H. G. Katzgraber, Phys. Rev. Lett. {\bf 101}, 197205 (2008).
\bibitem{banos:12} R. A. Ba\~{n}os, L. A. Fern\'andez, V. Mart\'in-Mayor, and A. P. Young, Phys. Rev. B {\bf 86}, 134416 (2012).
\bibitem{daboul:04} D. Daboul, I. Chang and A. Aharony, Eur. Phys. J. B {\bf 41}, 231 (2004).
\bibitem{lundow} P. H. Lundow and I. A. Campbell, arXiv:1402.1991.
\bibitem{ballesteros:99} H. G. Ballesteros, L. A. Fern\'andez, V. Mart\'in-Mayor, A. Mu\~noz Sudupe, G. Parisi, and J. J. Ruiz-Lorenzo, J. Phys. A {\bf 32} 1 (1999).
\bibitem{dedominicis} C. de Dominicis, private communication (2004).
\bibitem{bray} A. J. Bray, private communication (2004).
\bibitem{ballesteros:97} H. G. Ballesteros, L. A. Fern\'andez, V. Mart\'in-Mayor, A. Mu\~noz Sudupe, Phys. Letts. B {\bf 400}, 346 (1997).
\bibitem{klein:91} L. Klein, J. Adler, A. Aharony, A.B. Harris, and Y. Meir, Phys. Rev. B {\bf 43}, 11249 (1991).
\end{thebibliography}
\end{document}